\documentstyle[12pt]{article}
\begin{document}

\begin{titlepage}
\hskip 12cm \vbox{\hbox{BUDKERINP/97-86}\hbox{CS-TH
2/97}\hbox{October 1997}}
\vskip 0.3cm
\centerline{\bf QUARK-ANTIQUARK CONTRIBUTION TO THE BFKL
KERNEL$^{~\ast}$}
\vskip 0.8cm
\centerline{  V.S. Fadin$^{a,b~\dagger}$, R. Fiore$^{c,d~\ddagger}$,
A. Flachi$^{c~\ddagger}$, M.I. Kotsky$^{a~\dagger}$}
\vskip .3cm
\centerline{\sl $^{a}$ Budker Institute for Nuclear Physics,}
\centerline{\sl $^{b}$ Novosibirsk State University, 630090
Novosibirsk,
Russia}
\centerline{\sl $^{c}$ Dipartimento di Fisica, Universit\`a della
Calabria,}
\centerline{\sl $^{d}$ Istituto Nazionale di Fisica Nucleare, Gruppo
collegato di Cosenza,}
\centerline{\sl Arcavacata di Rende, I-87036 Cosenza, Italy}
\vskip 1cm
\begin{abstract}
The quark-antiquark contribution to the BFKL kernel is calculated.
Using  the effective vertex for the $q\bar q$ pair production in the
Reggeon-Reggeon collision we find this contribution by integrating 
the
square of this vertex over relative transverse momenta and fractions
of
longitudinal momenta of produced particles.
\end{abstract}
\vskip .5cm
\hrule
\vskip.3cm
\noindent

\noindent
$^{\ast}${\it Work supported in part by the Ministero italiano
dell'Universit\`a e della Ricerca Scientifica e Tecnologica, in part
by INTAS and in part by the Russian Fund of Basic Researches.}
\vfill
$ \begin{array}{ll}
^{\dagger}\mbox{{\it email address:}} &
 \mbox{FADIN, KOTSKY~@INP.NSK.SU}\\
\end{array}
$

$ \begin{array}{ll}
^{\ddagger}\mbox{{\it email address:}} &
  \mbox{FIORE, FLACHI~@FIS.UNICAL.IT}
\end{array}
$
\vfill
\vskip .1cm
\vfill
\end{titlepage}
\eject
\textheight 210mm \topmargin 2mm \baselineskip=24pt

\vskip 0.5cm

{\bf 1. Introduction}

The investigation of parton distributions in the region of small
values of
the Bjorken variable $x$ is nowadays one of the most important
problems of
perturbative QCD \cite{BCM}, especially in connection with results of
recent
experiments on deep inelastic scattering \cite{H1}. The calculation
of these
distributions in the leading logarithmic approximation (LLA), where
only
terms of the type $\alpha _s^n\ln ^n(1/x)$ are summed up to all
orders in $%
\alpha _s$, can be performed using the BFKL evolution equation
\cite{FKL}.
Now the results of the LLA are widely known and used for the analysis 
of the
experiments on semihard processes \cite{AKMS}. But to make the
analysis
reliable one has to know the radiative corrections to the leading
logarithmic results.

The BFKL equation can be presented in the form
\begin{equation}
\frac \partial {\partial ln(1/x)}{\cal F}(x,{\vec q_1^{~2}}) =
\int d^{D-2}q_2{\cal K}(\vec q_1,\vec q_2){\cal F}(x,{\vec 
q_2^{~2}})~,
\label{l1}
\end{equation}
where $D=4+2\epsilon~$ is the space-time dimension, different from 
$4$ to
have on
each step of calculations well defined expressions free from infrared
and
collinear divergencies, and the function ${\cal F}(x,\vec k^2)$ is 
the
unintegrated gluon density, connected with the gluon distribution
$g(x,Q^2)$
as follows:
\begin{equation}
xg(x,Q^2) = \int_0^{Q^2}d\left( {\vec k}^2\right) {\cal F}(x,{\vec 
k}^2)~.
\label{l2}
\end{equation}
Here $\vec k$ denotes the projection of the parton 4-momenta $k$ into
the
plane orthogonal to the initial particle momenta.The kernel ${\cal
K}(\vec
q_1,\vec q_2)$ is expressed in terms of the gluon Regge trajectory
$\omega
(t)\ $and the contribution of the real particle production
\begin{equation}
{\cal K}(\vec q_1,\vec q_2) =
2\omega (-{\vec q}_1^{~2})\delta (\vec q_1-\vec q_2)+
{\cal K}_{real}(\vec q_1,\vec q_2)~.
\label{l3}
\end{equation}

It was argued in Ref.~\cite{LF} that in the next-to-leading
logarithmic
approximation (NLLA) the form (\ref{l1}) of the BFKL equation, as 
well
as the
representation (\ref{l3}) of the kernel remain unchanged, but the
gluon
trajectory has to be taken in the two-loop approximation, whereas the
contribution of the real particle production is the sum of the
contributions coming from one-gluon, two-gluon and quark-antiquark 
pair
productions in the Reggeon-Reggeon collision:
\begin{equation}
{\cal K}_{real}(\vec q_1,\vec q_2) =
{\cal K}_{RRG}^{one-loop}(\vec q_1,\vec q_2)+
{\cal K}_{RRGG}^{Born}(\vec q_1,\vec q_2)+
{\cal K}_{RRQ\overline{Q}}^{Born}(\vec q_1,\vec q_2)~.
\label{l4}
\end{equation}
Let us remind that in the LLA only the first term in Eq.~(\ref{l4})
taken in
the Born approximation does contribute. In the NLLA this term has to
be calculated with the one-loop accuracy, in contrast with the second 
two
terms,
which are pure next-to-leading corrections.

The gluon Regge trajectory $\omega (t)$ was calculated in the
two-loop
approximation in Ref.~\cite{FFQ}. One-gluon production amplitudes
were found
with the required accuracy in Ref.~\cite{FL} and the contribution 
${\cal
K}%
_{RRGG}^{Born}$ for two-gluon production was obtained in
Refs.~\cite{FL96, FKL1}. The quark-antiquark production in the 
central
rapidity region was
considered in Refs.~\cite{CC, FL96, CC1}. The singular at $(\vec
q_1-\vec
q_2)^2\rightarrow 0$ part of ${\cal K}_{RRQ\overline{Q}}^{Born}(\vec
q_1,\vec q_2)$ was found for the case of massless quarks in
Ref.~\cite{FL96}; for this case in Ref.~\cite{CC1} the azimuthal 
averaged
regular part of this piece of the kernel is shown.

In this paper we calculate a complete expression for the massless
quark-antiquark contribution ${\cal K}_{RRQ\overline{Q}}^{Born}$ to
the BFKL
kernel. We use the representation of the amplitude of the $q\bar q$
pair
production in the central rapidity region in terms of the effective
vertex
for the $q\bar q$ pair production in the Reggeon-Reggeon collision.
The
expression for ${\cal K}_{RRQ\overline{Q}}^{Born}(\vec q_1,\vec q_2)$
is
obtained by integrating the square of this vertex over relative 
momenta
of
produced particles.

\vskip 0.5cm

{\bf 2. The quark-antiquark production in Reggeon-Reggeon collision}

The production of a quark-antiquark pair contributes to cross 
sections of
high
energy processes in the NLLA if the pair is produced in the
quasi multi-Regge
kinematics (QMRK), i.e. if it has a limited invariant mass and is 
divided
in the
rapidity space by large, growing with the energy intervals from other
produced
particles \cite{LF}. In this case the amplitudes of the processes 
contain
the
effective vertex $\gamma _{i_1i_2}^{Q{\overline{Q}}}(q_1,q_2)$ for 
the
quark-antiquark
production in the collision of two Reggeized gluons, having momenta
$q_1=\beta
p_A+q_{1\perp }$ and $-q_2=\alpha p_B-q_{2\perp }$ with $\alpha
,\beta \ll 1$
and colour indices $i_1,i_2.$ Here and below $\ p_A$ and $p_B$ are
light
cone vectors such that the momenta of the initial particles $A$ and 
$B$ are
equal to
$p_A+(m_A^2/s)p_B$ and $p_B+(m_B^2/s)p_A$, with $s=(p_A+p_B)^2.$
The effective vertex $\gamma _{i_1i_2}^{Q{\overline{Q}}}(q_1,q_2)$ 
can be
extracted
from any you like amplitude of the $q\bar q$ production in the QMRK. 
In
the
simplest case of the process $A+B\rightarrow A^{\prime }+B^{\prime
}+q\bar q$
we have

\begin{equation}
A_{AB}^{A^{\prime }Q{\overline{Q}}B^{\prime }} =
2s{\frac 1{\vec q_1^{~2}}}
\Gamma _{AA^{\prime }}^{i_1}\gamma _{i_1i_2}^{Q{\overline{Q}}}
(q_1,q_2){\frac1{\vec q_2^{~2}}}\Gamma _{BB^{\prime }}^{i_2}~,
\label{l5}
\end{equation}
where $q_1 = p_A-p_{A^{\prime }}$, $q_2=p_{B^{\prime }}-p_B$,
$\Gamma _{PP^{\prime}}^i$ are the particle-particle-Reggeon vertices 
at the
lowest order,
\begin{equation}
\Gamma _{PP^{\prime }}^i = g\langle P^{\prime }|T^i|P\rangle
\delta _{\lambda_P,\lambda _{P^{\prime }}}~;
\label{l6}
\end{equation}
here $\langle P^{\prime }|T^i|P\rangle $ represents matrix elements 
of the
group
generators, $\lambda _P$ are helicities of corresponding particles,
$g$ is
the gauge coupling constant ($\alpha _s=g^2/(4\pi )$). Using the 
results
of Ref.~\cite{FL96} we obtain
\begin{equation}
\gamma _{i_1i_2}^{Q{\overline{Q}}}(q_1,q_2) =
\frac 12g^2\bar u\left(k_1\right) \left[t^{i_1}t^{i_2}b(k_1,k_2)\,-
t^{i_2}\,t^{i_1}\overline{b(k_2,k_1)}\right] v(k_2)~,
\label{l7}
\end{equation}
where $t^i$ are the colour group generators in the fundamental
representation. We will use the Sudakov parametrization
\[
k_i = \beta _ip_A+\alpha _ip_B+k_{i\perp }~,~~~~~~
s\alpha _i\beta_i = -k_{i\perp }^2~ = \vec k_i^{~2},~~~~~~i=1,2~,
\]
\begin{equation}
\beta _1+\beta _2 = \beta \ll 1~,~~~~~~\alpha _1+\alpha _2 =
\alpha \ \ll 1~,
\label{l8}
\end{equation}
and the denotations
\[
x = \frac{\beta _1}{\beta _1+\beta _2}~,~~~~~~\Delta = 
k_1+k_2=q_1-q_2~,
\]
\begin{equation}
\Lambda = k_1-x\Delta ~,~~~~~~
Z = -\vec \Lambda ^{~2}-x(1-x)\vec \Delta^{~2}~.
\label{l9}
\end{equation}
The expressions for $b(k_1,k_2)$ and $\overline{b(k_2,k_1)}$ can be
presented in the following way:
\begin{equation}
b(k_1,k_2) = \frac{4\mbox{${\not{\hbox{\kern-2.0pt$p$}}}$}_A
\mbox{${\not{\hbox{\kern-2.0pt$Q$}}}$}_1
\mbox{${\not{\hbox{\kern-2.0pt$p$}}}$}_B}{st}-
\frac 1{{\Delta }^2}\mbox{${\not{\hbox{\kern-2.0pt$\Gamma$}}}$}
\label{l10}
\end{equation}
and
\begin{equation}
\overline{b(k_2,k_1)} = 
\frac{4\mbox{${\not{\hbox{\kern-2.0pt$p$}}}$}_B
\mbox{${\not{\hbox{\kern-2.0pt$Q$}}}$}_2
\mbox{${\not{\hbox{\kern-2.0pt$p$}}}$}_A}{s~\tilde t}-
\frac 1{{\Delta }^2}\mbox{${\not{\hbox{\kern-2.0pt$\Gamma$}}}$}~,
\label{l11}
\end{equation}
where
\[
t = (q_1-k_1)^2~,~~~~~~\tilde t = (q_1-k_2)^2~,
\]
\[
Q_1 = q_{1\perp }-k_{1\perp }~,~~~~~~Q_2 = q_{1\perp }-k_{2\perp }~,
\]
\begin{equation}
\Gamma = 2\left[ (q_1+q_2)_{\perp }-
\beta p_A\left(1+2x(1-x)\frac{\vec q_1^{~2}}Z\right)-
\frac{p_B}{\beta s}\left( 2\vec q_2^{~2}+\frac{Z}{x(1-x)}\right) 
\right]~.
\label{l12}
\end{equation}

The quark-antiquark contribution to the BFKL kernel can be put
\cite{FL96, FKL1} in the form
\[
{\cal K}_{RRQ\overline{Q}}^{Born}(\vec q_1,\vec q_2) =
\frac 1{2\vec q_1^{~2}\vec q_2^{~2}}\frac 1{(N^2-1)}\sum_{i_1,i_2,f}
\int d\kappa d\rho_f
\]
\begin{equation}
\times \delta ^{(D)}\,(q_1-q_2-k_1-k_2)
|\gamma_{i_1i_2}^{Q\overline{Q}}(q_1,q_2)|^2~,
\label{l13}
\end{equation}
where $k_1$ and $k_2$ are the quark and antiquark momenta, the sum is
taken over
the colour indices $i_1$, $i_2$ and over spin, colour and flavour
states of
the produced quark- antiquark pair, $\kappa =(q_1-q_2)^2$ is the
squared
invariant mass of the two Reggeons and the element $d\rho _f$ of the
phase
space is
\begin{equation}
d\rho _f = \prod_{n=1,2}\frac{d^{D-1}k_n}{(2\pi )^{D-1}2\omega _n}~.
\label{l14}
\end{equation}
From the representation (\ref{l7}) we obtain

\begin{equation}
\sum_{i_1,i_2,f}|\gamma_{i_1i_2}^{Q{\overline{Q}}}(q_1,q_2)|^2 =
\frac{g^4(N^2-1)n_f}{16N}[(N^2-1)A+B+(k_1\leftrightarrow k_2)]~,
\label{l15}
\end{equation}
where $n_f$ is the number of light quark flavours and

\begin{equation}
A = tr\left( \mbox{${\not{\hbox{\kern-2.0pt$k$}}}$}_1b(k_1,k_2)
\mbox{${\not{\hbox{\kern-2.0pt$k$}}}$}_2\overline{b(k_1,k_2)}\right)
\label{l16}
\end{equation}
and
\begin{equation}
B = tr\left( \mbox{${\not{\hbox{\kern-2.0pt$k$}}}$}_1b(k_1,k_2)
\mbox{${\not{\hbox{\kern-2.0pt$k$}}}$}_2b(k_2,k_1)\right) ~.
\label{l17}
\end{equation}
The calculation of the traces gives us
\[
A = 32x(1-x)
\left\{\vbox to 20.66pt{}-(1-x)(1-2x)^2\left( \vec q_1^{~2}\right)^2
\left( \frac{1}{xt^2}+\frac x{Z^2}\right)-
x(1-x)\frac{\vec q_1^{~2}\vec q_2^{~2}}{\vec \Lambda^2Z}\right.
\]
\[
\left. -4x^2(1-x)^2\frac{\vec q_1^{~2}(\vec \Lambda \vec \Delta )}
{\vec \Lambda^2Z}\left[ 2\frac{(\vec \Lambda \vec q_1)}{\vec 
\Lambda^2}+
x(1-x)\frac{(\vec \Lambda \vec \Delta )\vec q_1^{~2}}{\vec 
\Lambda^2Z}+
(1-2x)\frac{\vec q_1^{~2}}Z\right] \right.
\]
\[
\left. -4x(1-x)
\left[\vbox to 20.66pt{}\left( \frac{(\vec \Lambda \vec q_1)}{\vec 
\Lambda^2}+
\frac{\left( \vec q_1(\vec k_1-x\vec q_1)\right) 
}{xt}\right)^2\right.\right.
\]
\[
\left. \left. +\frac{\vec q_1^{~2}(\vec \Lambda \vec \Delta )}
{\vec \Lambda^2Zt}\left( (\vec k_1\vec q_1)+(1-2x)(\vec k_1\vec 
\Delta )+
x\vec \Delta^2-(\vec q_1\vec \Delta )\right) \right] \right.
\]
\[
\left. -(1-2x)\frac{\vec q_1^{~2}}t\left[ \frac{\vec q_2^{~2}}Z+
2(1-x)\frac{2(\vec k_1\vec q_1)-\vec q_1^{~2}}Z-
4(1-x)\frac{\vec q_1(\vec k_1-x\vec q_1)}{xt}\right] \right.
\]
\begin{equation}
\left. +2(1-x)(1-2x)\frac{\vec q_1^{~2}\left( (\vec \Lambda \vec 
\Delta)+
2(\vec \Lambda \vec q_1)\right) }{\vec \Lambda ^2}
\left( \frac 1t-\frac{x}{Z}\right) \right\}~,
\label{l18}
\end{equation}
and
\[
B-A+(k_1\leftrightarrow k_2) = 64x(1-x)
\left\{\vbox to 23.66pt{}\frac{1}
{2x(1-x)t\widetilde{t}}\left[-\vec q_1^{~2}\vec q_2^{~2}\right. 
\right.
\]
\[
\left. +2x(1-x)\vec q_1^{~2}(\vec q_2^{~2}-\vec \Delta ^2)+
8x(1-x)(\vec k_1\vec q_1)(\vec k_2\vec q_1)\right]
\]
\begin{equation}
\left. +(1-x)\frac{\left( \vec q_1^{~2}-2(\vec k_1\vec q_1)\right)^2}
{2xt^2}+x\frac{\left( \vec q_1^{~2}-2(\vec k_2\vec q_1)\right)^2}
{2(1-x)\widetilde{t}^2}\right\}~.
\label{l19}
\end{equation}

\vskip 0.5cm

{\bf 3. The quark-antiquark contribution to the BFKL kernel}

Going in Eq.~(\ref{l13}) to the variables $x,$ $\vec k_1$, we get

\begin{equation}
\int d\kappa d\rho _f\delta
^{(D)}\,(q_1-q_2-k_1-k_2) = \int_0^1\frac{dx}{2x(1-x)}\int 
\frac{d^{D-2}k_1}
{(2\pi )^{2(D-1)}}
\label{l20}
\end{equation}
and using Eq.~(\ref{l15}) we obtain
\[
{\cal K}_{RRQ\overline{Q}}^{Born}(\vec q_1,\vec q_2) =
\frac{g^4\mu^{2\epsilon }}{64\vec q_1^{~2}\vec q_2^{~2}(2\pi )^{D-1}}
\frac{n_f}{N}
\]
\begin{equation}
\times \int_0^1\frac{dx}{x(1-x)}\int 
\mu^{-2\epsilon}\frac{d^{D-2}k_1}
{(2\pi )^{D-1}}[(N^2-1)A+B+(k_1\leftrightarrow k_2)]~,
\label{l21}
\end{equation}
where $\mu $ is the renormalization scale. In order to obtain the
contribution to the BFKL kernel we should perform the integration in
Eq.~(\ref{l21}). The integrals appearing here have a similar
structure as
the integrals considered in Ref.~\cite{FKL1} for the calculation of 
the
two
gluon contribution to the BFKL kernel. The details of the calculation 
of
integrals for the quark-antiquark case will be given elsewhere
\cite{FFFK}. Here
we present only the results:
\begin{displaymath}
\int_0^1\frac{dx}{x(1-x)}\int \mu^{-2\epsilon }
\frac{d^{D-2}k_1}{(2\pi)^{D-1}}A =
\end{displaymath}
\begin{displaymath}
\frac{64\Gamma (1-\epsilon )}{(4\pi )^{2+\epsilon }}
\left\{\vbox to 16.66pt{} \frac{{\vec q_1}^{~2}{\vec q_2}^{~2}}{\vec 
\Delta^2}
\left( \frac{{\vec \Delta }^2}{\mu ^2}\right)^\epsilon
\frac{4\Gamma^2(2+\epsilon )}{\epsilon \Gamma(4+2\epsilon )} \right.
\end{displaymath}
\begin{displaymath}
\left. +\frac{{\vec q_1}^{~2}{\vec q_2}^{~2}}{{\vec q_1}^{~2}-{\vec 
q_2}^{~2}}
\left[ 1-\frac{{\vec \Delta }^2({\vec q_1}^{~2}+{\vec q_2}^{~2}+
4\vec q_1\vec q_2)}{3({\vec q_1}^{~2}-{\vec q_2}^{~2})^2}\right]
\ln \left( \frac{{\vec q_1}^{~2}}{{\vec q_2}^{~2}} \right) \right.
\end{displaymath}
\begin{equation}
\left. +\frac{{\vec \Delta }^2}{({\vec q_1}^{~2}-{\vec q_2}^{~2})^2}
\left( 2{\vec q_1}^{~2}{\vec q_2}^{~2}-\frac{1}{3}{\vec \Delta }^2
({\vec q_1}^{~2}+{\vec q_2}^{~2})\right)+
\frac{1}{3}\left( 2{\vec \Delta }^2-{\vec q_1}^{~2}-
{\vec q_2}^{~2}\right) \right\}~,
\label{l22}
\end{equation}
\[
\int_0^1\frac{dx}{x(1-x)}\int \mu ^{-2\epsilon }
\frac{d^{D-2}k_1}{(2\pi)^{D-1}}\left( B-A+(k_1\leftrightarrow 
k_2)\right) =
\]
\[
\frac{128\Gamma (1-\epsilon )}{(4\pi )^{2+\epsilon }}
\Biggl\{ \left[ -{\vec q_1}^{~2}{\vec q_2}^{~2}-\frac{({\vec 
q_1}^{~2}-
{\vec q_2}^{~2})^2}{4}+
\left( ({\vec q_1}{\vec q_2})^2-2{\vec q_1}^{~2}{\vec 
q_2}^{~2}\right)
\right.
\]
\[
\left. \times \frac{(2{\vec q_1}^{~2}{\vec q_2}^{~2}-3{\vec 
q_1}^{~4}-
3{\vec q_2}^{~4})}{216{\vec q_1}^{~2}{\vec q_2}^{~2}}\right]
\int_0^\infty \frac{dx}{({\vec q_1}^{~2}+x^2{\vec q_2}^{~2})}\ln
\left| \frac{1+x}{1-x}\right|
\]
\begin{equation}
+\frac{\left( 3({\vec q_1}{\vec q_2})^2-2{\vec q_1}^{~2}
{\vec q_2}^{~2}\right) }{16{\vec q_1}^{~2}{\vec q_2}^{~2}}
\left[ ({\vec q_1}^{~2}-{\vec q_2}^{~2})\ln \left( \frac{{\vec 
q_1}^{~2}}
{{\vec q_2}^{~2}}\right) +2({\vec q_1}^{~2}+{\vec 
q_2}^{~2})\right]\Biggr\}~.
\label{l23}
\end{equation}
Using these last equations, from Eq.~(\ref{l21}) we get the final
result for
the quark-antiquark contribution to the BFKL kernel:
\[
{\cal K}_{RRQ\overline{Q}}^{Born}(\vec q_1,\vec q_2) =
\frac{{4\bar g}_\mu^4\mu^{-2\epsilon }n_f}{\pi^{1+\epsilon }
\Gamma (1-\epsilon )N^3}
\Biggl\{ N^2\Biggl[ \frac 1{{\vec \Delta }^2}\left( \frac{{\vec 
\Delta}^2}
{\mu ^2}\right)^\epsilon \frac{2}{3}\left( \frac 1\epsilon 
-\frac{5}{3}+
\epsilon \left( \frac{28}9-\frac{\pi ^2}6\right) \right)
\]
\[
+\frac{1}{{\vec q_1}^{~2}-{\vec q_2}^{~2}}\left( 1-
\frac{{\vec \Delta }^2({\vec q_1}^{~2}+{\vec q_2}^{~2}+4{\vec q_1\vec 
q_2)}}
{3({\vec q_1}^{~2}-{\vec q_2}^{~2})^2}\right) \ln
\left( \frac{{\vec q_1}^{~2}}{{\vec q_2}^{~2}}\right)
\]
\[
+\frac{{\vec \Delta }^2}{({\vec q_1}^{~2}-{\vec q_2}^{~2})^2}\left( 
2-
\frac{{\vec \Delta }^2({\vec q_1}^{~2}+{\vec q_2}^{~2})}{3{\vec 
q_1}^{~2}
{\vec q_2}^{~2}}\right) +\frac{(2{\vec \Delta }^2-{\vec q_1}^{~2}-
{\vec q_2}^{~2})}{3{\vec q_1}^{~2}{\vec q_2}^{~2}}\Biggr]
\]
\[
+\Biggl[ -1-\frac{({\vec q_1}^{~2}-{\vec q_2}^{~2})^2}{4{\vec 
q_1}^{~2}
{\vec q_2}^{~2}}-\left( 2-\frac{({\vec q_1}{\vec q_2})^2}{{\vec 
q_1}^{~2}
{\vec q_2}^{~2}}\right) \frac{(2{\vec q_1}^{~2}{\vec q_2}^{~2}-
3{\vec q_1}^{~4}-3{\vec q_2}^{~4})}{16{\vec q_1}^{~2}{\vec 
q_2}^{~2}}\Biggr]
\]
\[
\times \int_0^\infty \frac{dx}{({\vec q_1}^{~2}+x^2{\vec q_2}^{~2})}
\ln \left| \frac{1+x}{1-x}\right| +\frac{\left( 3({\vec q_1}{\vec 
q_2})^2-
2{\vec q_1}^{~2}{\vec q_2}^{~2}\right) }{16{\vec q_1}^{~4}{\vec 
q_2}^{~4}}
\]
\begin{equation}
\times \left[ ({\vec q_1}^{~2}-{\vec q_2}^{~2})\ln
\left( \frac{{\vec q_1}^{~2}}{{\vec q_2}^{~2}}\right)+
2({\vec q_1}^{~2}+{\vec q_2}^{~2})\right]\Biggr\}~,
\label{l24}
\end{equation}
where
\begin{equation}
{\bar g}_\mu ^2 = \frac{g_\mu^2N\Gamma (1-{\epsilon })}{(4\pi 
)^{2+\epsilon }}
~,~~~~~~g_\mu =g{\mu }^\epsilon~.
\label{l25}
\end{equation}

\vskip 0.5cm

{\bf 4. Discussion}

The expression given in Eq.~(\ref{l24}) represents the correction to
the
BFKL kernel connected with the quark-antiquark production. It
includes all
terms which give nonvanishing contributions after integration over
$\vec
\Delta $ when $\epsilon $ $=(D-4)/2$ tends to its physical value
$\epsilon $
$=0.$ This correction contains two types of infrared singularities:
the
explicit pole in $\epsilon $ at fixed $\vec \Delta $ and the
singularity at $%
\vec \Delta ^2=0,$ which leads to a new pole in $\epsilon $ after
integration over $\vec \Delta $ . All singularities of
${\cal K}_{RRQ\overline{Q}}^{Born}$ are contained in the first line 
of
Eq.~(\ref{l24})
and agree with the corresponding result of Ref.~\cite{FL96}. The 
first
singularity is cancelled with the corresponding one in the virtual
correction to
the one-gluon production contribution; the singularities appearing
after
integration over $\vec \Delta $ cancel the corresponding
singularities in
the gluon Regge trajectory.

Let us define the singular part of ${\cal 
K}_{RRQ\overline{Q}}^{Born}$ as
given by the first line of Eq.~(\ref{l24}). It can be rewritten with
the required accuracy as
\begin{equation}
{\cal K}_{RRQ\overline{Q}}^{Born}(\vec q_1,\vec q_2)_{sing} =
\frac{16{\bar g}_\mu ^4\mu ^{-2\epsilon }n_f}
{\pi ^{1+\epsilon }\Gamma (1-\epsilon )N^{}}\frac{1}{{\vec \Delta 
}^2}
\left( \frac{{\vec \Delta }^2}{\mu^2}\right)^\epsilon
\frac{\Gamma ^2(2+\epsilon )}{\epsilon \Gamma (4+2\epsilon )}~.
\label{l26}
\end{equation}
In the remaining part of Eq.~(\ref{l24}) we can put $\epsilon $
$=0.$
Various terms in this part have singularities at
${\vec q_1^{~2}}=0,{\vec q_2^{~2}=0}$ and ${\vec q_1^{~2}=\vec 
q_2^{~2}}$.
But all these
singularities are spurious and cancel one another. The singularities 
at
${\vec q_1^{~2}=\vec q_2^{~2}}$ explicitly cancel after azimuthal 
averaging:
\[
<{\cal K}_{RRQ\overline{Q}}^{Born}(\vec q_1,\vec q_2)_{nonsing}> =
\frac{\alpha _s^2n_f}{4\pi^3N}
\left\{ \frac{2N^2}{3({\vec q_1}^{~2}-{\vec q_2}^{~2})}
\ln\left( \frac{{\vec q_1}^{~2}}{{\vec q_2}^{~2}} \right) \right.
\]
\[
\left. -\frac 1{32{\vec q_1}^{~2}{\vec q_2}^{~2}}\left[ ({\vec 
q_1}^{~2}-
{\vec q_2}^{~2})\ln \left( \frac{{\vec q_1}^{~2}}{{\vec 
q_2}^{~2}}\right)+
2({\vec q_1}^{~2}+{\vec q_2}^{~2})\right. \right.
\]
\begin{equation}
\left. \left. +(22{\vec q_1}^{~2}{\vec q_2}^{~2}-{\vec q_1}^{~4}-
{\vec q_2}^{~4})\frac{1}{\sqrt{{\vec q_1}^{~2}
{\vec q_2}^{~2}}}\biggl( \ln\left( \frac{{\vec q_1}^{~2}}{{\vec 
q_2}^{~2}}
\right)\mbox{arctg}\left( \frac{|{\vec q_2}|}{|{\vec q_1}|} \right) + 
2
\mbox{Im}~\mbox{Li}_2\left( i\frac{|{\vec q_2}|}{|{\vec q_1}|} 
\right)
\biggr) \right] \right\},
\label{l27a}
\end{equation}
where we have taken into account that
$$
\int_0^\infty \frac{dx}{({\vec q_1}^{~2}+x^2{\vec q_2}^{~2})}
\ln \left| \frac{1+x}{1-x}\right| =
$$
\begin{equation}\label{l27b}
= \frac{1}{\sqrt{{\vec q_1}^{~2}
{\vec q_2}^{~2}}}\biggl( \ln\left( \frac{{\vec q_1}^{~2}}{{\vec 
q_2}^{~2}}
\right)\mbox{arctg}\left( \frac{|{\vec q_2}|}{|{\vec q_1}|} \right) + 
2
\mbox{Im}~\mbox{Li}_2\left( i\frac{|{\vec q_2}|}{|{\vec q_1}|} 
\right)
\biggr).
\end{equation}
The expression (\ref{l27a}) should be compared with the result
given in
Eq.~(3.17) of Ref.~\cite{CC1}, but there are some misprints in
this equation, as it was indicated in Ref.~\cite{CC2}. Unfortunately,
corresponding expression of Ref.~\cite{CC2} is not free from 
missprints also.
To agree Eq.~(3.17) of
Ref.~\cite{CC1} with our
expression the first colour structure there should have an overall
1/2 factor and the factor 2 of the last term in square brackets 
should
multiply only the dilogarithm and not the
$\ln \frac 1\rho {\cal L}_1(\rho )$
piece; in the second term $C_A$ has to be changed for $n_f.$ We thank
M.Ciafaloni, who informed us that these are just the
misprints
in Eq.~(3.17) of Ref.~\cite{CC1}.

{\bf Acknowledgment}

One of us (V.S.F.) thanks the Dipartimento di Fisica dell'Universita
della
Calabria and the Istituto Nazionale di Fisica Nucleare - Gruppo
collegato di
Cosenza for their warm hospitality while a part of this work was
done.
\[
\]
\[
\]

\end{document}